\documentclass
[superscriptaddress,secnumarabic,amssymb,amsmath,nobibnotes,aps,prd,showkeys,showpacs,nofootinbib,onecolumn]{revtex4}%
\usepackage{graphicx}
\usepackage{epstopdf}
\usepackage{epsf}
\usepackage{bm}
\usepackage{amsmath}
\usepackage{amsfonts}
\usepackage{amssymb}%
\providecommand{\U}[1]{\protect\rule{.1in}{.1in}}

\newcommand{\be}{\begin{equation}}
\newcommand{\ee}{\end{equation}}

\newcommand{\mincir}{\raise
-3.truept\hbox{\rlap{\hbox{$\sim$}}\raise4.truept\hbox{$<$}\ }}
\newcommand{\magcir}{\raise
-3.truept\hbox{\rlap{\hbox{$\sim$}}\raise4.truept\hbox{$>$}\ }}

\begin{document}
\title{Dynamics of Einstein-Aether Scalar field Cosmology}
\author{Andronikos Paliathanasis}
\email{anpaliat@phys.uoa.gr}
\affiliation{Institute of Systems Science, Durban University of Technology, Durban 4000,
Republic of South Africa}
\author{G. Papagiannopoulos}
\email{yiannis.papayiannopoulos@gmail.com}
\affiliation{Faculty of Physics, Department of Astronomy-Astrophysics-Mechanics University
of Athens, Panepistemiopolis, Athens 157 83, Greece}
\author{Spyros Basilakos}
\email{svasil@academyofathens.gr}
\affiliation{Academy of Athens, Research Center for Astronomy and Applied Mathematics,
Soranou Efesiou 4, 11527, Athens, Greece}
\affiliation{National Observatory of Athens, Lofos Nymphon - Thissio, PO Box 20048 - 11810, Athens, Greece}
\author{John D. Barrow}
\email{J.D.Barrow@damtp.cam.ac.uk}
\affiliation{DAMTP, Centre for Mathematical Sciences, University of Cambridge, Wilberforce
Rd., Cambridge CB3 0WA, UK}

\begin{abstract}
We study the cosmological evolution of the field equations in the context of
Einstein-Aether cosmology by including a scalar field in a spatially flat
Friedmann--Lema\^{\i}tre--Robertson--Walker spacetime. Our analysis is
separated into two separate where a pressureless fluid source is included or
absent. In particular, we determine the critical points of the field equations
and we study the stability of the specific solutions. The limit of general
relativity is fully recovered, while the dynamical system admits de Sitter
solutions which can describe the past inflationary era and the future
late-time attractor. Results for generic scalar field potentials are presented
while some numerical behaviours are given for specific potential forms.

\end{abstract}
\date{\today}
\maketitle

\section{Introduction}

Einstein-Aether theory is a Lorentz-violating theory in which a unitary
timelike vector field, called the \ae ther, is introduced into the
Einstein-Hilbert action \cite{DJ,DJ2,Carru,carroll}. The introduction of the
timelike vector field in the action integral is also a specific selection of
preferred frame at each point in the spacetime, and so this modification
spontaneously breaks the Lorentz symmetry \cite{ea1}. The gravitational field
equations are of second-order and correspond to variations of the action with
respect to the metric tensor and the \ae ther field. At this point we recall
that the unitarity of the timelike vector field is guaranteed by introducing a
lagrange multiplier. The Einstein-Aether theory can describe various
cosmological phases, including those of early inflationary expansion and late
dark-energy domination\cite{data1,data2,data3,data4}. It is important to
mention here that the Einstein-Aether approach also describes the classical
limit of Ho\v{r}ava gravity \cite{esf}.

One of the ways to study a cosmological model is to perform a dynamical
analysis by studying its critical points in order to connect them to the
different observed eras, with their respective dynamical behaviours and
characteristics
\cite{DynSystemsWain,DynSystemsColey,Heinzle,dyn1,dyn2,dyn3,dyn4,dyn5,dyn6,dyn7,dyn9}%
. For the Einstein-Aether cosmologies there have been several such studies
\cite{dyn8,dyn8a,dyn08b,dyn08c,dyn08e,dyn08e,coleysf,KSEAPF}.

For Einstein-Aether cosmologies \cite{Barrow} provided exact solutions for
specific forms of the scalar field potential in the framework of
Friedmann--Lema\^{\i}tre--Robertson--Walker (FLRW) spacetime.
There has been further study of the dynamical evolution and stability of those
inflationary solutions in homogeneous and isotropic Einstein-Aether
cosmologies containing a self interacting scalar field which interacts with
the aether \cite{coleysf}.
Similar dynamical analysis can by found in \cite{Alhulaimi},
where it was shown that for isotropic expansion the dynamics are independent
of the aether parameters, but this is not the case for anisotropic expansion.
In all cases there is a period of slow-roll inflation at intermediate times
and, in some cases, accelerated expansion at late times.

Apart from the FLRW background scenario, there have been more wideranging
studies. One such work, investigating the dynamical equations of the
Einstein-Aether theory for the cases of FLRW as well as in an locally
rotationally symmetric Bianchi Type III geometry \cite{LangRoum}. There, it
was found that the existence or the non-existence of the solutions to the
reduced equations depends on the values of combinations of the initial
parameters that enter the action integral. Results of this type have also been
found elsewhere\ \cite{Foster,Jacobson,Zlosnik}. For other dynamical studies
in the context of the Einstein-Aether scenario we refer the reader the
articles of \cite{KSEAPF,ColeyLeon} \cite{eda5,eda6,eda7,eda8}.

The plan of this paper is as follows. In Section \ref{field} we present the
model to be studied, which is an Einstein-Aether scalar field cosmology with
spatially flat FLRW spacetime, where the scalar field lagrangian has been
modified so that the scalar field potential is non-minimally coupled to the
aether field, as proposed in \cite{DJ}. The dimensionless dynamical analysis
and the corresponding critical points are presented in Section \ref{sec3}. The
absence of the matter in the action integral implies that the dimension of the
dynamical system can be either one or two, while adding a pressureless fluid
raises the dimension of the system to two or three. Furthermore, the critical
points are classified into three families. Sections \ref{sec4} and \ref{sec5}
include the main results of the current analysis, where we present the allowed
critical points, It is interesting to mention that
the case of general relativity (GR) with a minimally coupled scalar field is
fully recovered, while new critical points are found which describe either
power-law or de Sitter solutions. Finally, we draw our conclusions in Section
\ref{sec6}.

\section{Einstein-aether cosmology}

\label{field}

First, we consider a spatially flat FLRW spacetime with line element%
\begin{equation}
ds^{2}=-dt^{2}+a^{2}\left(  t\right)  \left(  dr^{2}+r^{2}d\theta^{2}%
+r^{2}\sin^{2}\theta d\phi^{2}\right)  , \label{fr0a}%
\end{equation}
and as an aether field we consider the timelike vector field $u^{\mu}%
=\delta_{t}^{\mu}$. In Einstein-Aether Models the gravitational action is
given by \cite{jacobo1}%
\begin{equation}
S_{EA}=\int\sqrt{-g}dx^{4}\left(  R+\frac{c_{\theta}}{3}\theta^{2}+c_{\sigma
}\sigma^{2}+c_{\omega}\omega^{2}+c_{\alpha}\alpha^{2}\right)  +S_{\phi},
\label{fr0b}%
\end{equation}
where the action integral of the scalar field is assumed to be \cite{DJ}%
\begin{equation}
S_{\phi}=\int\sqrt{-g}dx^{4}\left(  \frac{1}{2}g^{\mu\nu}\phi_{,\mu}\phi
_{,\nu}-V\left(  \theta,\phi\right)  \right)  . \label{fr0c}%
\end{equation}
Parameters $\theta,~\sigma,~\omega$ and $\alpha$ describe the kinematic
quantities of the unitary vector field $u^{\mu},$ and correspond to the volume
\ expansion rate, shear, vorticity and fluid acceleration. Following the
notations of \cite{DJ} for the line element (\ref{fr0a}), the field equations
are written as (with units $8\pi G=c=1$)%

\begin{equation}
\frac{1}{3}\theta^{2}=\frac{1}{2}\dot{\phi}^{2}+V-\theta V_{\theta},
\label{fr1}%
\end{equation}%
\begin{equation}
\frac{2}{3}\dot{\theta}=-\dot{\phi}^{2}-\dot{\theta}V_{\theta\theta}-\dot
{\phi}V_{\theta\phi}, \label{fr2}%
\end{equation}%
\begin{equation}
\ddot{\phi}+\theta\dot{\phi}+V_{\phi}=0. \label{KG}%
\end{equation}

Barrow \cite{Barrow} previously found that for
\begin{equation}
V\left(  \phi,\theta\right)  =V_{0}e^{-\lambda\theta}+%
{\displaystyle\sum\limits_{r=0}^{n}}
V_{r}\theta^{r}e^{\frac{r-2}{2}\lambda\phi}, \label{pot.01a}%
\end{equation}
where $V_{0},~V_{r}$ and $\lambda$ are constants, the field equations
(\ref{fr1})-(\ref{KG}) admit exact power-law solutions $\phi\left(  t\right)
=\frac{2}{\lambda}\ln t~$,~$\theta\left(  t\right)  =3Bt^{-1}$, i.e. $a\left(
t\right)  \simeq t^{B}$ where $B=B\left(  V_{0},V_{r},\lambda\right)  $. The
detailed dynamical analysis of (\ref{pot.01a}) was performed in \cite{coleysf}.

In this work we consider the potential function $V\left(  \phi,\theta\right)
$ to be
\begin{equation}
V\left(  \phi,\theta\right)  =U\left(  \phi\right)  +Y\left(  \phi\right)
\theta, \label{pot.01}%
\end{equation}
where $U\left(  \phi\right)  $ now denotes the scalar field potential and
$Y\left(  \phi\right)  $ is the coupling term between the scalar and aether
fields. It is straightforward to observe that for $U\left(  \phi\right)
=U_{0}e^{-\lambda\theta}$ and $Y\left(  \phi\right)  \simeq\sqrt{U\left(
\phi\right)  }$ potential (\ref{pot.01a}) is recovered for $n=1,V_{0}=0$.

With the aid of (\ref{pot.01}) the field equations (\ref{fr1})-(\ref{KG})
simplify to%

\begin{equation}
\frac{1}{3}\theta^{2}=\frac{1}{2}\dot{\phi}^{2}+U\left(  \phi\right)  ,
\label{pot.02}%
\end{equation}%
\begin{equation}
\frac{2}{3}\dot{\theta}+\frac{1}{3}\theta^{2}=-\frac{1}{2}\dot{\phi}%
^{2}+U\left(  \phi\right)  -\dot{\phi}Y_{,\phi}, \label{pot.03}%
\end{equation}%
\begin{equation}
\ddot{\phi}+\theta\dot{\phi}+U_{,\phi}+Y_{,\phi}\theta=0. \label{pot.04}%
\end{equation}
from which we can rewrite them as
\begin{equation}
G_{ab}=T_{ab}, \label{pot.05}%
\end{equation}
where $G_{ab}$ is the Einstein tensor, and $T_{ab}$ describes the\ effective
Einstein-Aether fluid source with the scalar field, where in the $1+3$
decomposition is written as%
\begin{equation}
T_{ab}=\rho_{\phi}u_{a}u_{b}+p_{\phi}h_{ab}, \label{pot.06}%
\end{equation}
in which $h_{ab}=g_{ab}+u_{a}u_{b}$ is the projective tensor and $\rho_{\phi
}~$and $p_{\phi}$ are given by%
\begin{equation}
\rho_{\phi}=\frac{1}{2}\dot{\phi}^{2}+U\left(  \phi\right)  ~,~p_{\phi}%
=\frac{1}{2}\dot{\phi}^{2}-U\left(  \phi\right)  +\dot{\phi}Y_{,\phi}.
\label{pot.07}%
\end{equation}

We observe that for this specific potential, (\ref{pot.01}), the energy
density $\rho_{\phi}$ is defined as in Einstein's GR, while in the pressure
term $p_{\phi}$ a new part is introduced due to the coupling between the
scalar field and the aether field.

If a minimally coupled matter source is introduced with energy density
$\rho_{m}$, pressure $p_{m}$, and constant parameter for the equation of
state, $w_{m}=p_{m}/\rho_{m},$ the field equations (\ref{pot.02}%
)-(\ref{pot.03}) are modified as follows:%
\begin{equation}
\frac{1}{3}\theta^{2}=\frac{1}{2}\dot{\phi}^{2}+U\left(  \phi\right)
+\rho_{m} \label{pot.08}%
\end{equation}%
\begin{equation}
\frac{2}{3}\dot{\theta}+\frac{1}{3}\theta^{2}=-\frac{1}{2}\dot{\phi}%
^{2}+U\left(  \phi\right)  -\dot{\phi}Y_{,\phi}-w_{m}\rho_{m} \label{pot.09}%
\end{equation}
where, for the perfect fluid $\rho_{m}$, the conservation law is%
\begin{equation}
\dot{\rho}_{m}+\theta\left(  1+w_{m}\right)  \rho_{m}=0. \label{pot.10}%
\end{equation}

We carry out our analysis by writing the field equations (\ref{pot.08}%
)-(\ref{pot.10}) in dimensionless variables by using expansion-normalised variables.

\section{Dynamical system}

\label{sec3}

In this section we present the main features of the dynamical analysis by
using the method of critical points. This method is powerful because it
provides information concerning the general evolution of the dynamical system.
Hence, from such an analysis the overall cosmological viability of the model
can be discussed.

The new dimensionless variables are defined as follows \cite{copeland}%
\begin{equation}
x=\sqrt{\frac{3}{2}\frac{\dot{\phi}^{2}}{\theta^{2}}}~,~y=\sqrt{\frac
{3U}{\theta^{2}}}~,~\lambda=\frac{U_{,\phi}}{U}~,\xi=\sqrt{2}\frac{Y_{,\phi}%
}{\sqrt{U}}~,~\Omega_{m}=\frac{3\rho_{m}}{\theta^{2}}. \label{var}%
\end{equation}
After some calculations, the system of the field equations is written in these
variables as
\begin{equation}
\Omega_{m}=1-x^{2}-y^{2}, \label{pot.11}%
\end{equation}%
\begin{equation}
\frac{dx}{dN}=-3x(1+J\left(  x,y,\xi,\Omega_{m}\right)  )\ -\frac{1}{2}\left(
\sqrt{6}\lambda y+3\xi\right)  y, \label{pot.12}%
\end{equation}
{}%

\begin{equation}
\frac{dy}{dN}=\sqrt{\frac{3}{2}}\lambda xy-3yJ,\label{pot.13}%
\end{equation}%
\begin{equation}
\frac{d\lambda}{dN}=\sqrt{6}x\lambda^{2}(\Gamma_{\lambda}\left(
\lambda\right)  -1),\ \label{pot.14}%
\end{equation}%
\begin{equation}
\frac{d\xi}{dN}=\ \sqrt{3}\xi x\left(  \Gamma_{\xi}\left(  \xi\right)
-\frac{\lambda}{\sqrt{2}}\right)  ,\label{pot.15}%
\end{equation}
where $N=\ln\left(  a\right)  $, function~$J\left(  x,y,\xi,\Omega_{m}\right)
$ is expressed as%
\begin{equation}
2J\left(  x,y,\xi,\Omega_{m}\right)  =-1-x^{2}+y^{2}-\xi xy-w_{m}\Omega
_{m}.\label{pot.16}%
\end{equation}
and
\begin{equation}
\Gamma_{\lambda}\left(  \lambda\right)  =\frac{U_{,\phi\phi}U}{(U_{,\phi}%
)^{2}}~~,~\Gamma_{\xi}\left(  \xi\right)  =\sqrt{2}\frac{Y_{,\phi\phi}%
}{Y_{,\phi}}\label{pot.17}%
\end{equation}

In general, equations (\ref{pot.11})-(\ref{pot.15}) form a three-dimensional
system. Specifically, equations (\ref{pot.14}), (\ref{pot.15}) do not coexist,
because parameters $\lambda$ and $\xi$ are not independent. Locally, the
condition $\frac{\partial\lambda}{\partial\phi}\neq0$ implies that the inverse
function of $\lambda\left(  \phi\right)  $ exists so that $\phi=\phi\left(
\lambda\right)  $. Hence the function $\xi\left(  \phi\right)  $ depends on
$\lambda$; indeed, $\xi=\xi\left(  \phi\left(  \lambda\right)  \right)  $,
so~$\xi=\xi\left(  \lambda\right)  $. In that case the only independent
variables which survive are the $\left\{  x,y,\lambda\right\}  $.

On the other hand, if locally $\lambda=const$, then $\frac{d\lambda}{dN}%
\equiv0$, and $\xi=\xi\left(  \phi\right)  $; hence, the only independent
variables that survive are $\left\{  x,y,\xi\right\}  ,$ since $\phi
=\phi\left(  \xi\right)  $. Consequently, there are three large families of
potentials which we will study, that admit different dynamical systems:

\textbf{Family (A)} with $\lambda=const$. and $\xi=cons\not t ,$ which
corresponds to the potentials
\begin{equation}
U\left(  \phi\right)  =U_{0}e^{\lambda\phi}~,~Y\left(  \phi\right)
=Y_{0}+Y_{1}e^{-\frac{\lambda}{2}\phi}, \label{pot.18}%
\end{equation}

\textbf{Family (B)} where $U\left(  \phi\right)  =U_{0}e^{\lambda\phi}$,
$\lambda=const$ with $\xi=\xi\left(  \phi\right)  ,$ and

\textbf{Family (C)} where the potential $U\left(  \phi\right)  $ is different
from the exponential potential and so $\xi=\xi\left(  \lambda\right)  .$

We observe from equation (\ref{pot.11}) that the variables $x,y$ obey the
inequalities $0\leq x^{2}+y^{2}\leq1$, where in the special case of
$\Omega_{m}=0$, this reduces to $x^{2}+y^{2}=1.$ At this point we mention that
the equation of state parameter for the perfect fluid is now
\begin{equation}
w_{\phi}\left(  x,y,\lambda,\xi,\Omega_{m}\right)  =-1+\frac{2x^{2}+\xi
xy}{x^{2}+y^{2}}, \label{pot.19}%
\end{equation}
while the deceleration parameter is
\begin{equation}
q\left(  x,y,\lambda,\xi,\Omega_{m}\right)  =-1-2J, \label{pot.20}%
\end{equation}
and the equation of state parameter for the effective fluid is
\begin{equation}
w_{\mathrm{tot}}\left(  x,y,\lambda,\xi,\Omega_{m}\right)  =\frac{1}{3}\left(
q-1\right)  \text{.} \label{pot.21}%
\end{equation}
We continue our study with the analysis of the critical points of the system
(\ref{pot.11})-(\ref{pot.15}) for the aforementioned three families of potentials.


\section{Scalar field without matter source}

\label{sec4}

In this section we consider the case where the cosmic fluid does not include
matter, namely $\Omega_{m}=0$. In this case our results are summarized as follows.

\subsection{Family A}

For the first family of potentials the constraint (\ref{pot.11}) implies that
the dynamical system (\ref{pot.12})-(\ref{pot.13}) is reduced to a
one-dimensional system. In this case we derive four critical points $\left(
x_{P},y_{P}\right)  $:

a. Points $A_{1\left(  \pm\right)  }$ with coordinates $A_{1\left(
\pm\right)  }=\left(  \pm1,0\right)  $. These points describe a universe where
$\rho_{\phi}=p_{\phi}=\dot{\phi}^{2}/2$, hence $w_{\phi}=\frac{p_{\phi}}%
{\rho_{\phi}}=1$. For the stability of the points we need to calculate the
corresponding eigenvalues, which in this case are given by $e_{1}\left(
A_{1\left(  \pm\right)  }\right)  =3\pm\sqrt{\frac{3}{2}}\lambda.$ Therefore,
point $A_{1\left(  +\right)  }$ is stable for $\lambda<-\sqrt{6}$ while point
$A_{1\left(  -\right)  }$ is stable for $\lambda>\sqrt{6}$.

b. Point $A_{2}$ with coordinates $A_{2}=\left(  -\frac{2\sqrt{2}\lambda
+\xi\sqrt{3\left(  4+\xi^{2}\right)  -2\lambda^{2}}}{\sqrt{3}\left(  4+\xi
^{2}\right)  },\frac{-\sqrt{2}\lambda\xi+2\sqrt{3\left(  4+\xi^{2}\right)
-2\lambda^{2}}}{\sqrt{3}\xi\left(  4+\xi^{2}\right)  }\right)  ~$exists for
$\left\{  \lambda<-\sqrt{6}~,~\xi\geq\sqrt{\frac{2\left(  \lambda
^{2}-6\right)  }{3}}\right\}  ~$or~$\left\{  \lambda>\sqrt{6}~,~\xi\leq
-\sqrt{\frac{2\left(  \lambda^{2}-6\right)  }{3}}\right\}  $ or~$\left\{
-\sqrt{6}\leq\lambda\leq\sqrt{6}~,~\xi\neq0\right\}  .$ The equation of state
parameter is written as
\begin{equation}
w_{\phi}\left(  \lambda,\xi\right)  =-1+\lambda\frac{4\lambda+\xi
\sqrt{6\left(  4+\xi^{2}\right)  -4\lambda^{2}}}{3\left(  4+\xi^{2}\right)
}.~
\end{equation}
Point $A_{2}$ describes an accelerated universe when $w_{\phi}<-\frac{1}{3}$;
that is, the parameters $\lambda,\xi$ are constrained by the following
conditions ~$\left\{  \lambda<-\sqrt{2}~,~\xi>\sqrt{2}\frac{\lambda^{2}%
-2}{\sqrt{3\lambda^{2}-2}}\right\}  ~$or ~$\left\{  \lambda>\frac{\sqrt{6}}%
{3}~,~0<\xi<-\sqrt{2}\frac{\lambda^{2}-2}{\sqrt{3\lambda^{2}-2}}\right\}  ~$or
$\left\{  -\sqrt{2}\leq\lambda\leq\frac{\sqrt{6}}{3}~,~\xi>0\right\}  $
or~$\left\{  \lambda\geq\sqrt{6}~,~\xi<-\sqrt{2}\frac{\lambda^{2}-2}%
{\sqrt{3\lambda^{2}-2}}\right\}  .$ As far as the stability is concerned we
find that $A_{2}$ is stable when $\left\{  0<\lambda\leq\sqrt{6}%
~,~\xi>0\right\}  ~$or~$\left\{  \lambda=\sqrt{6}~,~\xi<0\right\}  $ or
$\left\{  \lambda>\sqrt{6}~,~\xi\leq-\sqrt{\frac{2}{3}\left(  \lambda
^{2}-6\right)  }\right\}  .~$ Moreover, if $A_{2}$ is an attractor describing
cosmic acceleration, then the parameters $\lambda,\xi$ obey the inequalities
$\left\{  -\sqrt{2}<\lambda\leq\sqrt{\frac{2}{3}}~,~\xi>0\right\}
~$or~$\left\{  \lambda<-\sqrt{2}~,~0<\xi<\sqrt{2}\frac{\lambda^{2}-2}%
{\sqrt{3\lambda-2}}\right\}  $ \ or $\left\{  \lambda\geq\frac{\sqrt{6}~}%
{3},0<~\xi<-\sqrt{2}\frac{\lambda^{2}-2}{\sqrt{3\lambda-2}}\right\}  $ or
$\left\{  \lambda\geq\sqrt{6}~,~\xi<-\sqrt{2}\frac{\lambda^{2}-2}%
{\sqrt{3\lambda-2}}~\right\}  $

c. Point $A_{3}$\thinspace\thinspace$=\left(  \frac{-2\sqrt{6}\lambda
+\sqrt{9\xi^{4}-6\left(  \lambda^{2}-6\right)  \xi^{2}}}{3\left(  4+\xi
^{2}\right)  },\frac{-\sqrt{6}\xi^{2}\lambda+\sqrt{9\xi^{4}-6\left(
\lambda^{2}-6\right)  \xi^{2}}}{3\xi\left(  4+\xi^{2}\right)  }\right)  $
exists when~$\left\{  \lambda<-\sqrt{6}~,~\xi\geq\sqrt{6\left(  \lambda
^{2}-6\right)  }\right\}  $ or $\left\{  \lambda=-\sqrt{6}~,~\xi\neq0\right\}
~$\ or $\left\{  -\sqrt{6}<\lambda<\sqrt{6},~\xi<0\right\}  $ \ or $\left\{
\lambda>\sqrt{6}~,~\xi<-\frac{1}{3}\sqrt{6\left(  \lambda^{2}-6\right)
}\right\}  $. The parameter of the equation of state is
\begin{equation}
w_{\phi}\left(  \lambda,\xi\right)  =-1+\lambda\frac{4\lambda-\sqrt{6\xi
^{4}-4\left(  \lambda^{2}-6\right)  }}{3\left(  4+\xi^{2}\right)  },
\end{equation}
where $w_{\phi}<-\frac{1}{3}$ when $\left\{  -\sqrt{2}<\lambda<-\sqrt{\frac
{2}{3}},-\sqrt{\frac{2\left(  2-\lambda^{2}\right)  ^{2}}{3\lambda^{2}-2}}%
<\xi<0\right\}  $ or $\ \left\{  -\sqrt{\frac{2}{3}}\leq\lambda\leq\sqrt
{2}~,~\xi<0\right\}  $ or $\left\{  \lambda>\sqrt{2},~\xi<-\sqrt
{\frac{2\left(  2-\lambda^{2}\right)  ^{2}}{3\lambda^{2}-2}}\right\}  $.
\ Point $A_{3}$ describes a stable solution only for $\xi<0$ and more
specifically $\left\{  0<\lambda<\sqrt{6}~,~-\xi_{1}<\xi<0\right\}  $ or
$\left\{  -\sqrt{6}\leq\lambda\leq\sqrt{6}~,~\xi<0\right\}  $ or $\left\{
\lambda>\sqrt{6}~,~\xi_{2}<\xi<0\right\}  $, where $\xi_{1},\xi_{2}$ are the
solutions of the algebraic equation%
\begin{equation}
8\left(  6-\lambda^{2}\right)  -\left(  120+50\lambda^{2}\right)  \xi
^{2}+75\xi^{4}=0.
\end{equation}

\begin{figure}[ptb]
\textbf{ \includegraphics[width=0.4\textwidth]{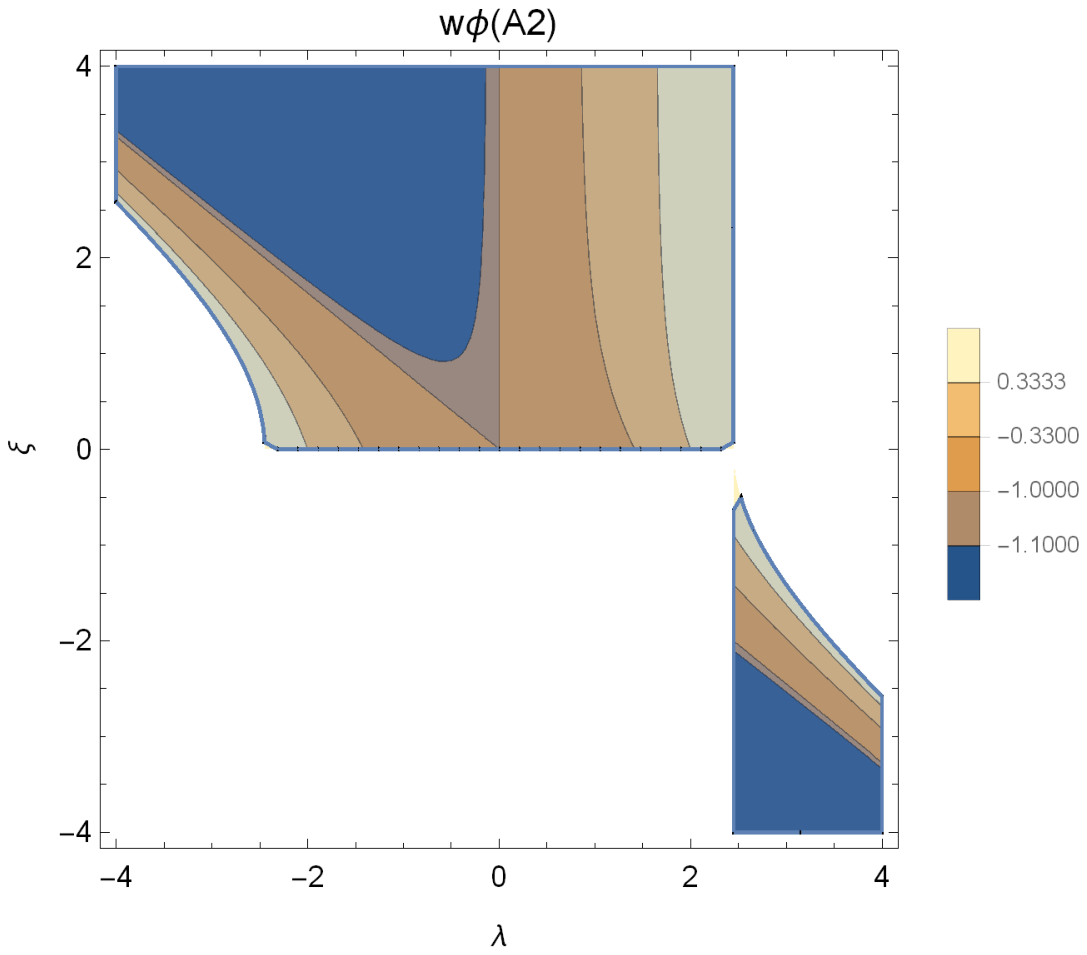}
\includegraphics[width=0.4\textwidth]{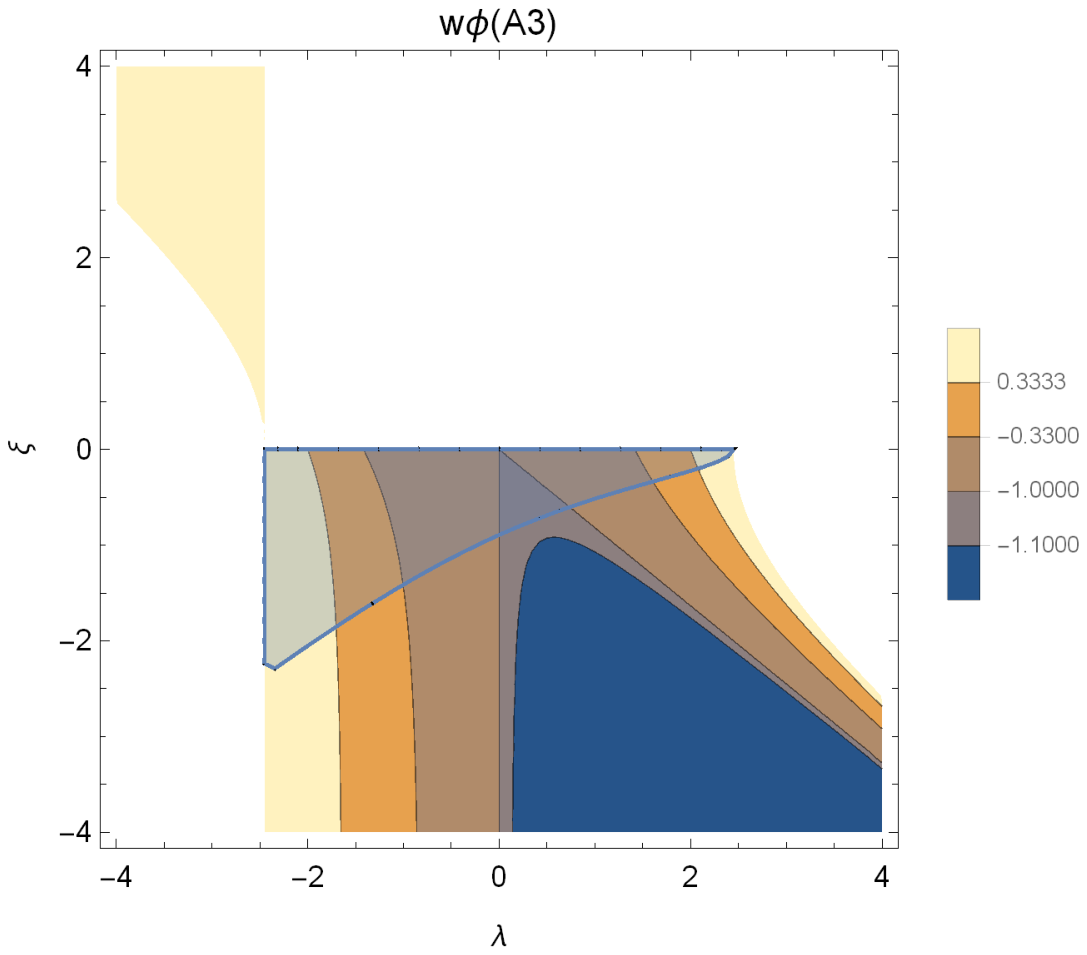} \newline}\caption{Region plot
in the space $\left\{  \lambda,\xi\right\}  $ for $w_{\phi}$ in the case of
critical points $A_{2}$ and $A_{3}$. The stable critical points are
represented by shaded regions. Shaded regions define the areas where the
critical points are stable. }%
\label{fig1}%
\end{figure}

In Fig. \ref{fig1} we present the contour plots of the equation of state
parameter $w_{\phi}\left(  \lambda,\xi\right)  $ for points $A_{2}$ and
$A_{3}$. Notice that the stable critical points are represented by shaded
regions.
We observe that points with $\lambda>0$ and $\xi<0$ describe stable
accelerated solutions, while it is possible for the EoS parameter to cross the
phantom line, namely $w_{\phi}<-1$.

\subsection{\textbf{Family B}}

We continue our analysis with the second family of critical points, namely
Family B. Here the dynamical system is formed by equations (\ref{pot.12}),
(\ref{pot.13}) and (\ref{pot.15}). By including the constraint, the dimension
of the system is reduced by one, i.e. from three to two dimensions. We study
the general evolution of the dynamical system by considering a general
function $\Gamma_{\xi}\left(  \xi\right)  $ \cite{fm1,fm3,fm4}.

The critical points $\left(  x_{p},y_{p},\xi_{p}\right)  ~$of the dynamical
system are:

a. Points $B_{1\left(  \pm\right)  }=\left(  A_{1\left(  \pm\right)  },\xi
_{0}\right)  $ for which $\xi_{0}$ is a solution of the algebraic equation
$\Gamma_{\xi}\left(  \xi_{0}\right)  =\frac{\lambda}{\sqrt{2}}~,~$or $\xi
_{0}=0.$ The points describe the same physical solution as those for
$A_{1\left(  \pm\right)  }$, hence the existence of the critical points is
given in section 4.1, however the stability conditions change Specifically,
the two eigenvalues are $e_{1}\left(  B_{1\left(  \pm\right)  }\right)
=6\pm\sqrt{6}\lambda$ and $e_{2}\left(  B_{1\left(  \pm\right)  }\right)
=\pm\sqrt{3}\xi_{0}\Gamma_{\xi}^{\prime}\left(  \xi_{0}\right)  .~$ Therefore,
$B_{1\left(  +\right)  }$ is stable when $\lambda<-\sqrt{6}~,~\xi_{0}%
\Gamma_{\xi}^{\prime}\left(  \xi_{0}\right)  <0$, while $B_{1\left(  -\right)
}$ is stable as long as $\lambda>\sqrt{6}~,~\xi_{0}\Gamma_{\xi}^{\prime
}\left(  \xi_{0}\right)  >0$.

b. Point $B_{2}=\left(  A_{2}\left(  \xi_{0}\right)  ,\xi_{0}\right)  $ with
$\Gamma_{\xi}\left(  \xi_{0}\right)  =\frac{\lambda}{\sqrt{2}}$. Again the
properties of $B_{2}$ are similar with those of $A_{2}$ (see section 4.1).
Concerning stability conditions $B_{2}$, describes an attractor solution when
$\xi_{0}\Gamma_{\xi}^{\prime}\left(  \xi_{0}\right)  >0:$ $\left\{
\lambda<0~,\xi_{0}>\frac{\sqrt{6}}{3}\left\vert \lambda\right\vert \right\}  $
or $\left\{  0<\lambda\leq\sqrt{6}~,~\xi_{0}>0\right\}  $.

c. Point $B_{3}=\left(  A_{3}\left(  \xi_{0}\right)  ,\xi_{0}\right)  $ with
$\Gamma_{\xi}\left(  \xi_{0}\right)  =\frac{\lambda}{\sqrt{2}}$. The physical
properties of $B_{3}$ are those of point $A_{3}$ (see previous section).
$B_{3}$ describes a stable solution for $\xi_{0}\Gamma_{\xi}^{\prime}\left(
\xi_{0}\right)  <0:\left\{  -\sqrt{6}<\lambda\leq0,~\xi_{0}<0\right\}  $ or
$\left\{  \lambda>0,\xi<-\frac{\sqrt{6}}{3}\lambda\right\}  $.

d. Point $B_{4}$ with coordinates $B_{4}=\left(  -\frac{\lambda}{\sqrt{6}%
},\sqrt{1-\frac{\lambda^{2}}{6}},0\right)  $ \ exists for $\left\vert
\lambda\right\vert <\sqrt{6}$. This situation describes a tracking solution of
the exponential potential with $\xi=0$. The equation of state parameter reads
$w_{\phi}\left(  \lambda,\xi\right)  =-1+\frac{\lambda^{2}}{3}$, hence we have
acceleration when $\left\vert \lambda\right\vert <\sqrt{2}$. The eigenvalues
of the linearized system are determined to be $e_{1}\left(  B_{4}\right)
=-3+\lambda^{2}~,~e_{2}\left(  B_{4}\right)  =\frac{\lambda}{2}\left(
\lambda-\sqrt{2}\Gamma_{\xi}\left(  0\right)  \right)  ,~$where $e_{1}\left(
B_{4}\right)  <0,\ e_{2}\left(  B_{4}\right)  <0$ for $\left\vert
\lambda\right\vert <\sqrt{3}$, and $\left\vert \Gamma_{\xi}\left(  0\right)
\right\vert >\frac{\sqrt{2}}{2}\lambda$.

e. Finally, point $B_{5}$ with coordinates $B_{5}=\left(  0,1,\xi_{0}\right)
,~\xi_{0}=-\sqrt{\frac{2}{3}}\lambda$ describes a de Sitter solution for which
$w_{\phi}=-1$. The eigenvalues of the linearized system are $e_{1}\left(
B_{5}\right)  =-\frac{3+\sqrt{3\left(  3-2\lambda^{2}+2\sqrt{2}\lambda
\Gamma_{\xi}\left(  \xi_{1}\right)  \right)  }}{2}~,~e_{1}\left(
B_{5}\right)  =-\frac{3-\sqrt{3\left(  3-2\lambda^{2}+2\sqrt{2}\lambda
\Gamma_{\xi}\left(  \xi_{1}\right)  \right)  }}{2}~$and thus point $B_{5}$ is
an attractor when $\left\vert \Gamma_{\xi}^{\prime}\left(  \xi_{0}\right)
\right\vert <\frac{\lambda}{\sqrt{2}}$. Notice that the de Sitter solution
does not exist for the exponential case in the context of the scalar field
cosmology which reduces to GR.

\subsubsection{Application}

Consider $Y\left(  \phi\right)  =Y_{1}e^{\left(  \nu+\frac{\lambda}{2}\right)
\phi},~\nu\neq0$; we calculate that $\phi=\frac{1}{2}\ln\left(  \frac{2\xi
^{2}}{Y_{1}^{2}\left(  2\nu+\lambda\right)  ^{2}}\right)  $ and $\Gamma\left(
\xi\right)  =\left(  \sqrt{2}v+\frac{\lambda}{\sqrt{2}}\right)  $.

Therefore, equation (\ref{pot.15}) is simplified to%
\begin{equation}
\frac{d\xi}{dN}=\ \sqrt{6}\nu\xi x,
\end{equation}
while the possible critical points are now only points $B_{1\left(
\pm\right)  }~$with $\xi_{0}=0$, $B_{4}$ and $B_{5}$. Points $B_{4}$ and
$B_{5}$ are attractors when $\left\{  \left\vert \lambda\right\vert <\sqrt
{6},\nu\lambda>0\right\}  $ and $\left\{  \nu<0~,~0<\lambda\leq-\frac{3}{2\nu
}\right\}  \cup\left\{  \nu>0~,~-\frac{3}{2\nu}<\lambda\leq0\right\}  ,$ respectively.

\begin{figure}[ptb]
\textbf{ \includegraphics[width=0.7\textwidth]{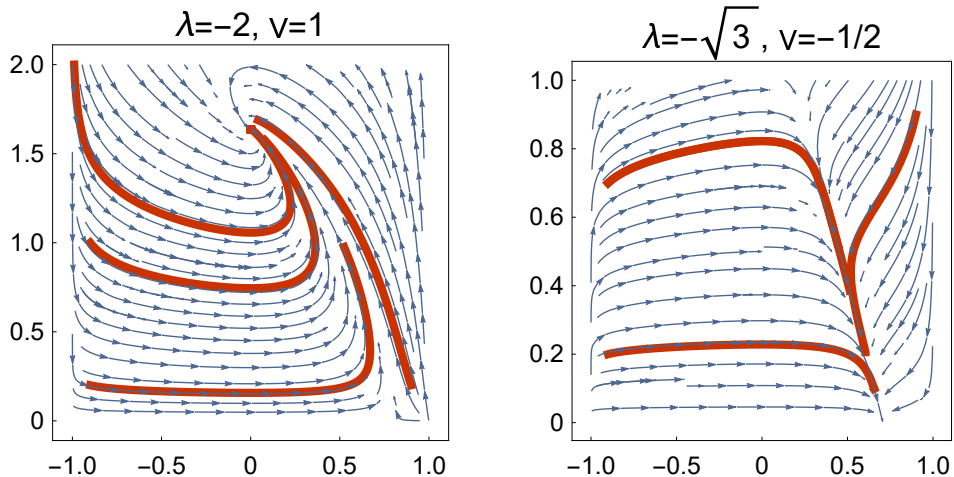} }\caption{Phase space
portrait in the variables $\left\{  x,\xi\right\}  $ for the dynamical system
(\ref{pot.12}), (\ref{pot.13}) and (\ref{pot.15}) $\ $and for $Y\left(
\phi\right)  =Y_{1}e^{\left(  \nu+\frac{\lambda}{2}\right)  \phi}~$without a
matter source. Left-hand figure is for $\lambda=-2$ and $\nu=1$ where we see
that point $B_{5}$ is the attractor of the system. $\backslash$Right-hand
figure is for $\lambda=-\sqrt{3}$~and $\nu=-\frac{1}{2}$ where the attractor
is point $B_{4}$ ,which describes the tracking solution. Solid lines describe
real trajectories}%
\label{fig2}%
\end{figure}

In Fig. \ref{fig2} we present the phase space diagram for the dynamical system
in the variables $\left\{  x,\xi\right\}  $ for two sets of the variables
$\lambda~$and $\nu$. For $\lambda=-2$ and $\nu=1$ it is clear that the de
Sitter universe $B_{5}~$is an attractor while for $\lambda=-\sqrt{3}$,
$\nu=-\frac{1}{2}$ the unique attractor is the scaling solution $B_{4}$.

In a similar way we continue with the third family of critical
points that we considered.

\subsection{Family C}

The third family of critical points correspond to the dynamical system
(\ref{pot.12}), (\ref{pot.13}) and (\ref{pot.14}) with the constraint
condition (\ref{pot.11}). Recall that in this case $\xi=\xi\left(
\lambda\right)  $.

\begin{figure}[ptb]
\textbf{ \includegraphics[width=0.6\textwidth]{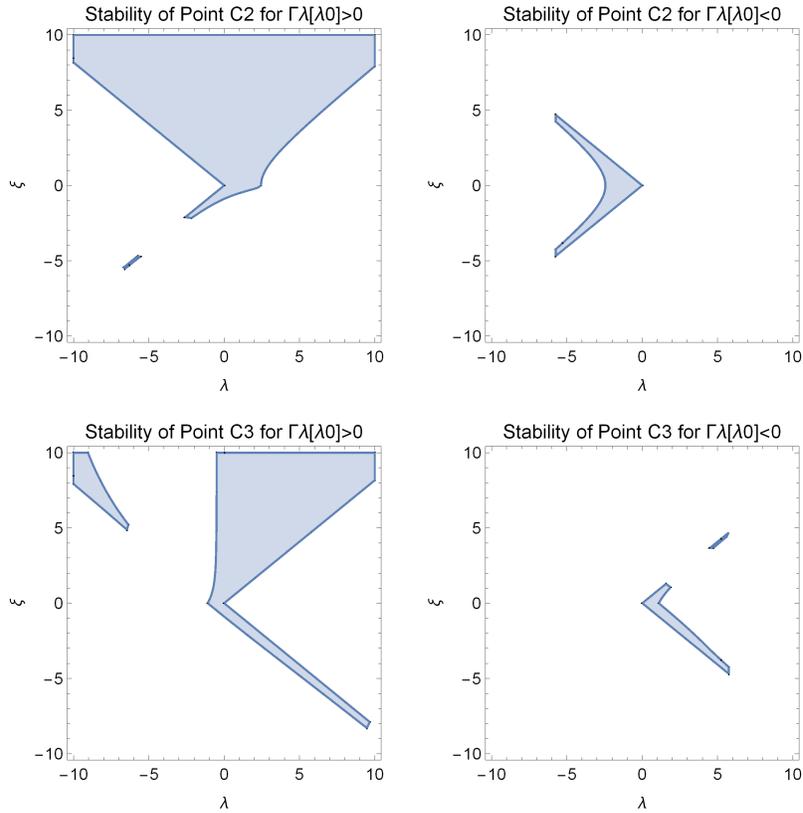} }\caption{Region
plots for the variables $\left\{  \lambda_{0},\xi\left(  \lambda_{0}\right)
\right\}  $ in which points $C_{2}$ and $C_{3}$ have eigenvalues with negative
real parts. Left-hand figures are for $\Gamma_{\lambda}^{\prime}\left(
\lambda_{0}\right)  >0,$ while right-hand figures are for $\Gamma_{\lambda
}^{\prime}\left(  \lambda_{0}\right)  \,<0$.}%
\label{fig4b}%
\end{figure}

The critical points are:

a. Points $C_{1\left(  \pm\right)  }=\left(  A_{1\left(  \pm\right)  }%
,\lambda_{0}\right)  $ with $\Gamma\left(  \lambda_{0}\right)  =1$ or
$\lambda_{0}=0.$ The physical descriptions of the points are those of
$A_{1\left(  \pm\right)  }$. The eigenvalues of the linearized system are
calculated to be $e_{1}\left(  C_{1\left(  \pm\right)  }\right)  =6-\sqrt
{6}\lambda_{0}~,~e_{2}\left(  C_{1\left(  \pm\right)  }\right)  =-\sqrt
{6}\lambda_{0}^{2}\Gamma^{\prime}\left(  \lambda_{0}\right)  ,$. We observe
that at least of one of the points $C_{1\left(  \pm\right)  }$ is stable only
for $\Gamma^{\prime}\left(  \lambda_{0}\right)  >0$ and $\left\vert
\lambda_{0}\right\vert >\sqrt{6}$.

b. Point $C_{2,3}=\left(  A_{2,3},\lambda_{0}\right)  $ with $\Gamma\left(
\lambda_{0}\right)  =1$. The existence conditions and the physical description
are the same as those of $A_{2,3}$ (see section 4.1). Because of the
nonlinearity of the eigenvalues, the map of $\lambda_{0},\xi\left(
\lambda_{0}\right)  $ in which the points are stable is presented in Fig.
\ref{fig4b}.


c. Point $C_{4}=\left(  -\frac{\xi}{\sqrt{4+\xi^{2}}},\frac{2}{\sqrt{4+\xi
^{2}}},\lambda_{0}\right)  $ with $\lambda_{0}=0,$ describes a de Sitter
solution, $w_{\phi}\left(  C_{4}\right)  =-1$, and actually reduces to points
$C_{2}$, $C_{3}$, with $\lambda_{0}=0$ respectively. The eigenvalues of the
linearized system are given by $e_{1}\left(  C_{4}\right)  =0~,~e_{2}\left(
C_{4}\right)  =-3.~$ Since $e_{1}=0$ we apply the central manifold theorem in
order to decide the stability and we find that $C_{4}$ is always an attractor.

d. Point $C_{5}=\left(  B_{5},\lambda_{0}\right)  $ with $\lambda_{0}%
=-\sqrt{\frac{3}{2}}\xi$ is found to be stable when the following condition
holds $\left(  \Gamma_{\lambda}\left(  0\right)  -1\right)  \left(  2+\sqrt
{6}\xi^{\prime}\left(  0\right)  \right)  >0\,$. The physical description of
$C_{5}$ is the same as that of $B_{5}$.


\subsubsection{Application}

\begin{figure}[ptb]
\textbf{ \includegraphics[width=0.7\textwidth]{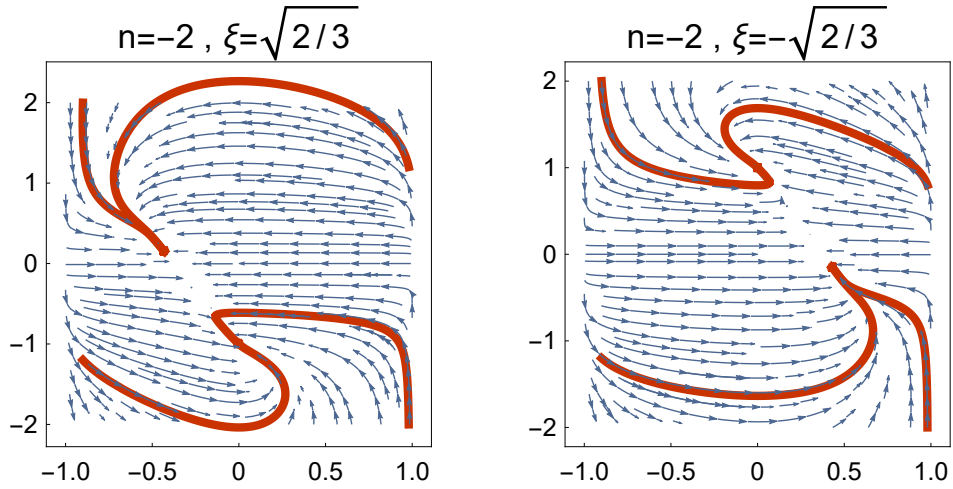} }\caption{Phase space
portrait for in the variables $\left\{  x,\lambda\right\}  $ for the dynamical
system (\ref{pot.12}), (\ref{pot.13}) and (\ref{pot.16}) $\ $and for $U\left(
\phi\right)  =U_{0}\phi^{n}$ and $Y\left(  \phi\right)  =Y_{0}\phi^{1+\frac
{2}{n}}~$without matter source. Left figure is for $n=-2$ and $\xi=\sqrt
{\frac{2}{3}}$ from where points $C_{4}$ and $C_{5}$ are attractor of the
system. Right figure is for $n=-2$ and $\xi=-\sqrt{\frac{2}{3}}$ where again
$C_{4}$ and $C_{5}$ are attractor of the system. Solid lines describe real
trajectories}%
\label{fig3aa}%
\end{figure}

\begin{figure}[ptb]
\textbf{ \includegraphics[width=0.7\textwidth]{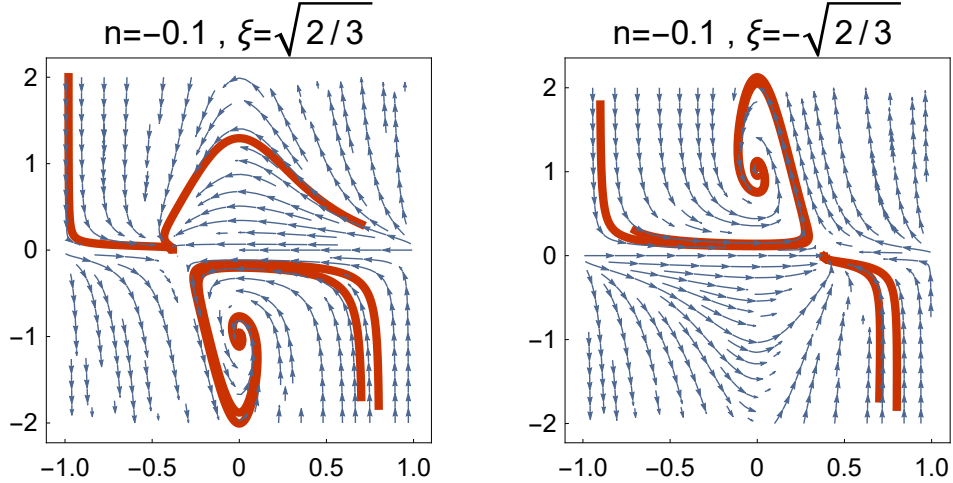} }\caption{Phase space
portrait for in the variables $\left\{  x,\lambda\right\}  $ for the dynamical
system (\ref{pot.12}), (\ref{pot.13}) and (\ref{pot.16}) $\ $and for $U\left(
\phi\right)  =U_{0}\phi^{n}$ and $Y\left(  \phi\right)  =Y_{0}\phi^{1+\frac
{2}{n}}$ without matter source. Left figure is for $n=-0.1$ and $\xi
=\sqrt{\frac{2}{3}}$ while right figure is for $n=-0.1$ and $\xi=-\sqrt
{\frac{2}{3}}$. It is clear that the only attractor is $C_{4}$ while $C_{5}$
is a source point. Solid lines describe real trajectories}%
\label{fig4aa}%
\end{figure}

Let us consider $U\left(  \phi\right)  =U_{0}\phi^{n}$ and $Y\left(
\phi\right)  =Y_{0}\phi^{1+\frac{2}{n}}$ from where we calculate $\phi
=\frac{n}{\lambda}$, $\xi=\frac{Y_{0}}{\sqrt{2U_{0}}}\left(  2+n\right)
=const$ and $\Gamma_{\lambda}\left(  \lambda\right)  =\frac{n-1}{n}$.
Therefore, equation (\ref{pot.14}) reduces to
\begin{equation}
\frac{d\lambda}{dN}=-\frac{\sqrt{6}}{n}x\lambda^{2}.
\end{equation}
Therefore, the critical points are $C_{1\left(  \pm\right)  },~C_{4}$ and
$C_{5}$. As far as stability is concerned we find that points $C_{1\left(
\pm\right)  }$ are always unstable and point $C_{5}~$is stable only when
$n\leq-2\xi^{2}$. For the latter case using various values of the free
parameters $n,~\xi$ we plot in Figs. \ref{fig3aa} and \ref{fig4aa} the phase
space diagram $\left\{  x,\lambda\right\}  $.


\section{Scalar field in the presence of matter}

\label{sec5}

In this section we include in our analysis a pressureless matter component
with $w_{m}=\frac{p_{m}}{\rho_{m}}=0$. In this case since $0\leq\Omega_{m}%
\leq1$, the constraint equation (\ref{pot.11}) yields $x^{2}+y^{2}\leq1$.
Following the lines of the previous section we study the same family of
potentials, namely A, B and C.

\subsection{Family A}

a. The first Point is $\bar{A}_{0}=\left(  0,0\right)  $ and the arbitrary
parameters $\lambda,~\xi$ describe a universe for which $\Omega_{m}=1$,
$w_{\mathrm{tot}}=0$. The eigenvalues of the linearized system are determined
to be $e_{1}\left(  \bar{A}_{0}\right)  =-\frac{3}{2}$,~$e_{2}\left(  \bar
{A}_{0}\right)  =\frac{3}{2}$, hence the point is always unstable.

b. Points $\bar{A}_{1\left(  \pm\right)  }$ with eigenvalues $e_{1}\left(
\bar{A}_{1\left(  \pm\right)  }\right)  =3\pm\sqrt{\frac{3}{2}}\lambda
~,~e_{2}\left(  \bar{A}_{1\left(  \pm\right)  }\right)  =3$, from which we
infer that points $\bar{A}_{1\left(  \pm\right)  }$ are unstable points.

c. Point $\bar{A}_{2}$ with eigenvalues $e_{1}\left(  \bar{A}_{2}\right)
=-3+\frac{4\lambda^{2}+\lambda\sqrt{6\xi^{4}-4\left(  \lambda^{2}-6\right)  }%
}{4+\xi^{2}}~,~e_{2}\left(  \bar{A}_{2}\right)  =-3+\frac{2\lambda^{2}%
+\lambda\sqrt{6\xi^{4}-4\left(  \lambda^{2}-6\right)  }}{2\left(  4+\xi
^{2}\right)  }$ describes a stable solution when $\left\{  \lambda<-\sqrt
{3}~,~\xi>\frac{2\left(  \lambda^{2}-3\right)  }{\sqrt{6\lambda^{2}-9}%
}\right\}  ~$or $\left\{  -\sqrt{3}\leq\lambda\leq\frac{\sqrt{6}}{2}%
~,~\xi>0\right\}  $ or $\left\{  \lambda>\frac{\sqrt{6}}{2}~,~\xi
<\frac{6-2\lambda^{2}}{\sqrt{6\lambda^{2}-9}}\right\}  $ or~$\left\{
\lambda=-\sqrt{6},~\xi>\frac{2\sqrt{3}}{3}\right\}  $.

d. Point $\bar{A}_{3}$ with eigenvalues $e_{1}\left(  \bar{A}_{3}\right)
=-3+\frac{4\lambda^{2}-\lambda\sqrt{6\xi^{4}-4\left(  \lambda^{2}-6\right)  }%
}{4+\xi^{2}}~,~e_{2}\left(  \bar{A}_{3}\right)  =-3+\frac{2\lambda^{2}%
-\lambda\sqrt{6\xi^{4}-4\left(  \lambda^{2}-6\right)  }}{2\left(  4+\xi
^{2}\right)  }$ describes a stable solution when $\left\{  \lambda
<-\frac{\sqrt{6}}{2}~,~\frac{2\left(  \lambda^{2}-3\right)  }{\sqrt
{6\lambda^{2}-9}}<\xi<0\right\}  ~$or $\left\{  -\frac{\sqrt{6}}{2}\leq
\lambda\leq\sqrt{3}~,~\xi<0\right\}  $ or $\left\{  \lambda>\sqrt{3}%
~,~\xi<-\frac{2\left(  \lambda^{2}-3\right)  }{\sqrt{6\lambda^{2}-9}}\right\}
$ or $\left\{  \lambda=\sqrt{6}~,~\xi<-\frac{2\sqrt{3}}{3}\right\}  $.

e. Point $\bar{A}_{4}$ with coordinates $\bar{A}_{4}=\left(  -\sqrt{\frac
{3}{2}}\frac{1}{\lambda},\sqrt{\frac{3}{2}}\frac{\sqrt{\lambda^{2}\left(
4+\xi^{2}\right)  }-\lambda\xi}{2\lambda^{2}}\right)  $ describes a universe
where the scalar field mimics the pressureless fluid, i.e. $w_{\phi}=0$. The
effective parameter is $w_{\mathrm{tot}}=0$, where $\Omega_{m}=1-\frac
{3}{\lambda^{2}}-\frac{3\xi^{2}}{4\lambda^{2}}+\frac{3}{4}\frac{\xi}%
{\lambda^{3}}\sqrt{\lambda^{2}\left(  4+\xi^{2}\right)  }$. The point exists
when $\left\{  \lambda<-\frac{\sqrt{6}}{2}~,~\xi\leq\frac{2\left(  \lambda
^{2}-3\right)  }{\sqrt{6\lambda^{2}-9}}\right\}  $~or$~\left\{  \sqrt{\frac
{3}{2}}<\lambda<\sqrt{3}~,~\xi\geq\frac{2\left(  \lambda^{2}-3\right)  }%
{\sqrt{6\lambda^{2}-9}}\right\}  $ or $\left\{  \lambda\geq\sqrt{3}~,~\xi
\geq-\frac{2\left(  \lambda^{2}-3\right)  }{\sqrt{6\lambda^{2}-9}}\right\}  $.

\begin{figure}[ptb]
\textbf{ \includegraphics[width=0.6\textwidth]{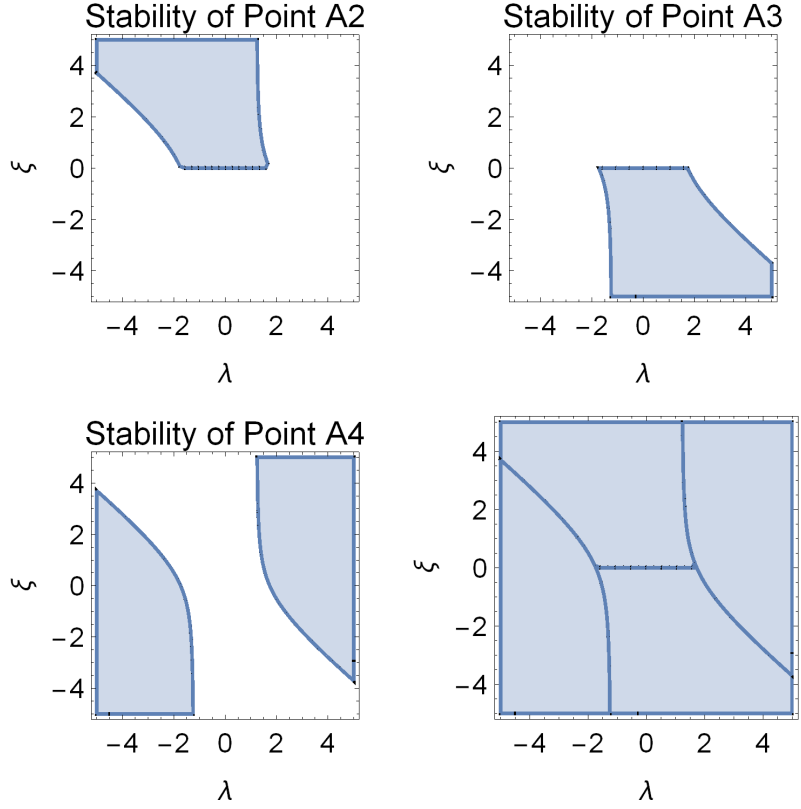} }\caption{Region
plots for the parameters $\left\{  \lambda,\xi\right\}  $ in which the
critical points $A_{2},~A_{3}$ and $A_{4}$ are stable. The lower right-hand
figure shows the common regions where we observe that there is not any common
intersection. Hence, there is only one possible stable point for the dynamical
system. }%
\label{fig5}%
\end{figure}

In Fig. \ref{fig5} we show $\left\{  \lambda,\xi\right\}  $ diagrams for where
the critical points $\bar{A}_{2}~,~\bar{A}_{3}$ and $\bar{A}_{4}$ exist and
have negative eigenvalues, namely we have stable solutions. Moreover, from
Fig. \ref{fig5} we observe that only one of the critical points $\bar{A}_{2}%
$,~$\bar{A}_{3}$ and $\bar{A}_{4}$ can be stable points of the system.

\begin{figure}[ptb]
\textbf{ \includegraphics[width=0.9\textwidth]{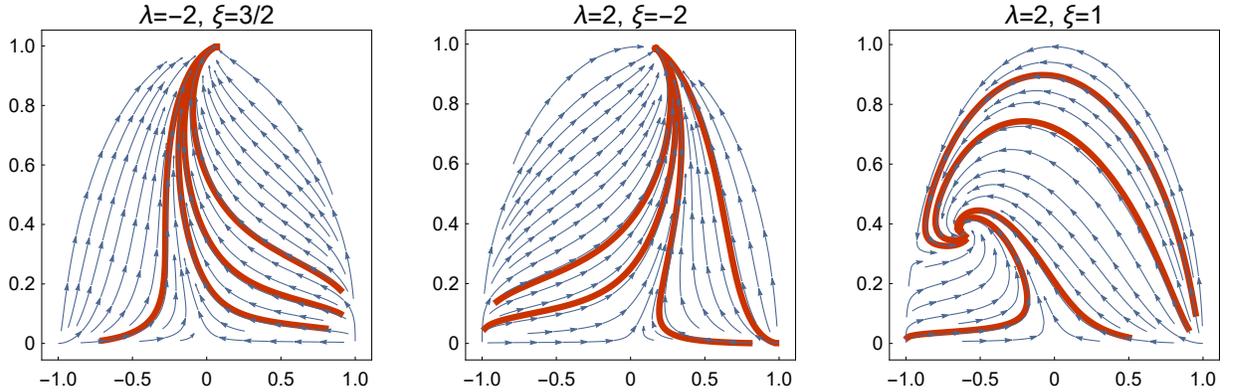} }\caption{Phase space
diagrams for the dynamical system (\ref{pot.12}), (\ref{pot.13}) with three
different sets of the free parameters $\lambda,\xi$ where one of the critical
points $\bar{A}_{2},~\bar{A}_{3}$ and $\bar{A}_{4}$ is stable. For $\left(
\lambda,\xi\right)  =\left(  -2,\frac{3}{2}\right)  $ point $A_{2}$ is the
unique attractor; for $\left(  \lambda,\xi\right)  =\left(  2,-2\right)  $
point $\bar{A}_{3}$ is the unique attractor, while point $\bar{A}_{4}$ is an
attractor for the plot with $\left(  \lambda,\xi\right)  =\left(  2,1\right)
$. Solid lines correspond to real trajectories.}%
\label{fig6}%
\end{figure}

\begin{figure}[ptb]
\textbf{ \includegraphics[width=0.6\textwidth]{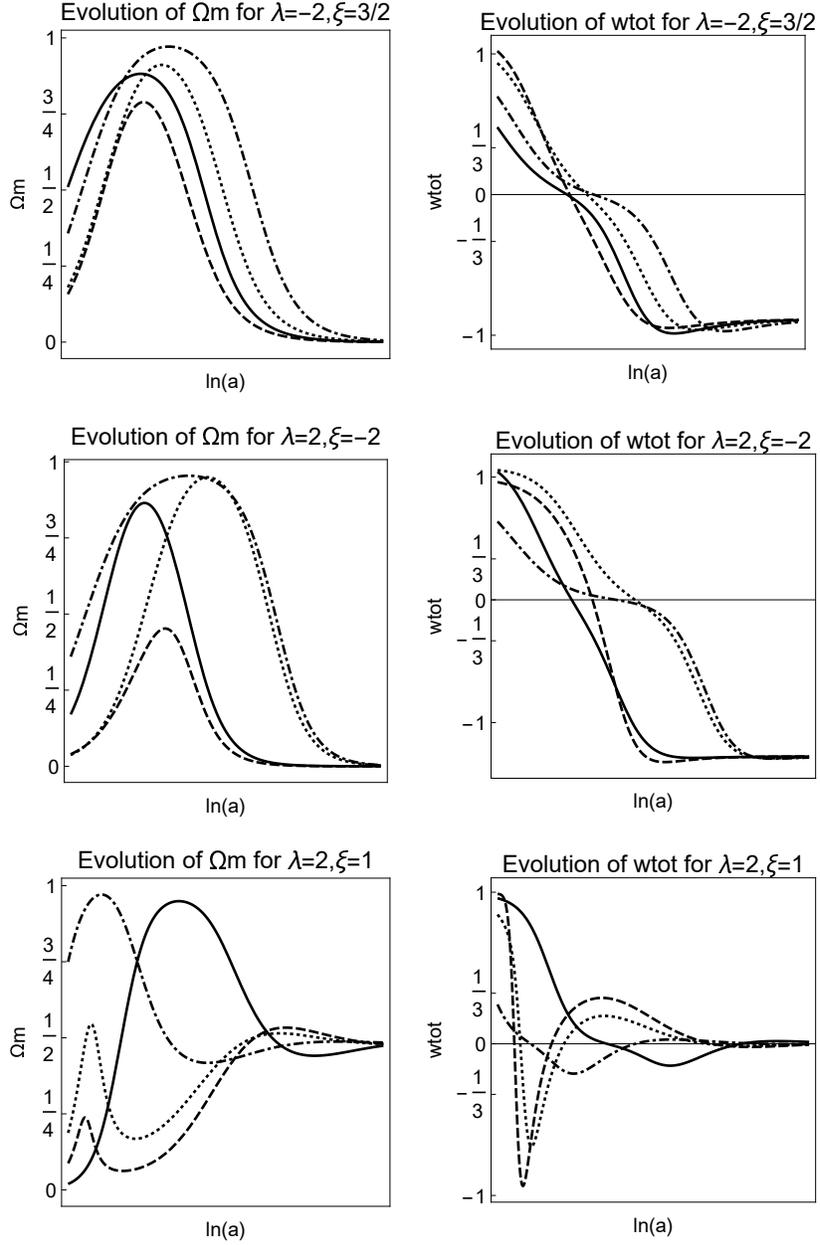} }\caption{Qualitative
evolution of the energy density $\Omega_{m}$ and of the parameter of the
equation of state for the effective fluid for the real trajectories presented
in Fig. \ref{fig6}. }%
\label{fig7}%
\end{figure}

In Fig. \ref{fig6} we present the phase space diagrams for the dynamical
variables $\left\{  x,y\right\}  $, using different values of the free
parameters $\lambda,\xi$. In particular, we provide three diagrams where in
each case one of the critical points $\bar{A}_{2},~\bar{A}_{3}$ and $\bar
{A}_{4}$ is stable. The dynamical evolution of the cosmological parameters
$\Omega_{m}\left(  a\right)  $ and $w_{tot}\left(  a\right)  $ are
demonstrated in Fig. \ref{fig7} for the real trajectories presented in Fig.
\ref{fig6}.

\subsection{Family B}

\begin{figure}[ptb]
\includegraphics[width=0.6\textwidth]{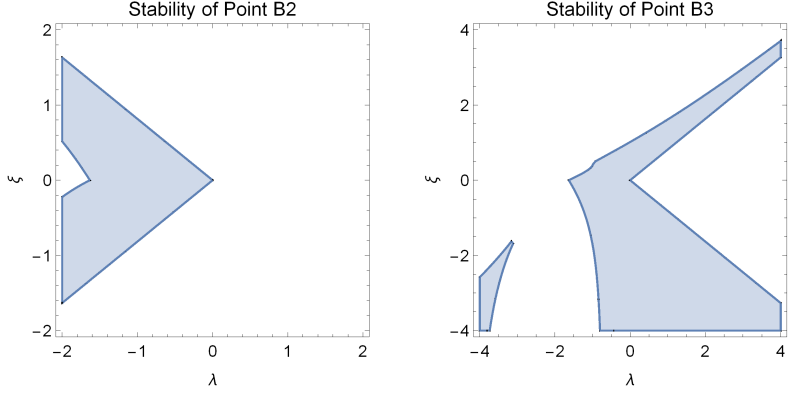} \caption{Region plots for the
variables $\left\{  \lambda,\xi\right\}  $ in which point $\bar{B}_{2}$
(left-hand figure) is an attractor and $\xi_{0}\Gamma_{\xi}^{\prime}\left(
\xi_{0}\right)  >0$ and point $\bar{B}_{3}$ (right-hand figure) describe a
stable solution with $\xi_{0}\Gamma_{\xi}^{\prime}\left(  \xi_{0}\right)  <0.$
}%
\label{fig8}%
\end{figure}

\begin{figure}[ptb]
\includegraphics[width=0.32\textwidth]{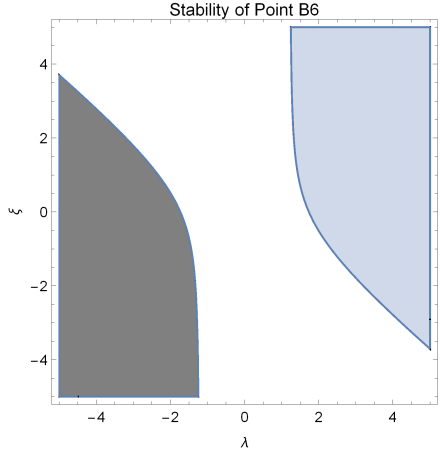} \caption{Region plots for the
variables $\left\{  \lambda,\xi\right\}  $ in which point $\bar{B}_{5}$ is an
attractor. The blue area is for $\xi_{0}\Gamma_{\xi}^{\prime}\left(  \xi
_{0}\right)  \geq0$ while the grey area is for $\xi_{0}\Gamma_{\xi}^{\prime
}\left(  \xi_{0}\right)  \leq0$. }%
\label{fig9}%
\end{figure}

The critical points which correspond to family B are the following:

a. Point~$\bar{B}_{0}=\left(  \bar{A}_{0},\xi\right)  $ where $\xi$ is
arbitrary: since $\xi$ is arbitrary, point $B_{0}$ describes a line in the
space $\left\{  x,y,\xi\right\}  $. The physical description is that of point
$A_{0}$. The eigenvalues of the linearized system are determined to be
$e_{1}\left(  \bar{B}_{0}\right)  =-\frac{3}{2}~,~e_{2}\left(  \bar{B}%
_{0}\right)  =\frac{3}{2}$ and $e_{3}\left(  \bar{B}_{0}\right)  =0$, from
which we conclude that $B_{0}$ describes an unstable solution and $B_{0}$ is a
source point.

b. Points $\bar{B}_{1\left(  \pm\right)  }$ with eigenvalues $e_{1}\left(
\bar{B}_{1\left(  \pm\right)  }\right)  =3~$,~$e_{2}\left(  \bar{B}_{1\left(
\pm\right)  }\right)  =3\pm\sqrt{6}\lambda~,$ $e_{3}\left(  \bar{B}_{1\left(
\pm\right)  }\right)  =\pm\sqrt{3}\xi_{0}\Gamma_{\xi}^{\prime}\left(  \xi
_{0}\right)  $, hence the solutions at points $B_{1\left(  \pm\right)  }$ are unstable.

c. Point $\bar{B}_{2}$ which is found to describe a stable solution if and
only if $\xi_{0}\Gamma_{\xi}^{\prime}\left(  \xi_{0}\right)  >0$. Notice that
the parameters $\left\{  \lambda,\xi_{0}\right\}  $ can be viewed in Fig.
\ref{fig8}.

d. Point $\bar{B}_{3}$ which is found to describe a stable solution if and
only if $\xi_{0}\Gamma_{\xi}^{\prime}\left(  \xi_{0}\right)  <0$, while the
parameters $\left\{  \lambda,\xi_{0}\right\}  $ are given in Fig. \ref{fig8}.

e. Point $\bar{B}_{4}$ with eigenvalues $e_{1}\left(  \bar{B}_{4}\right)
=-3+\frac{\lambda^{2}}{2}$,~$e_{2}\left(  \bar{B}_{4}\right)  =-3+\lambda^{2}$
and $e_{3}\left(  \bar{B}_{3}\right)  =\frac{\lambda^{2}}{2}-\frac{\sqrt
{2}\lambda}{2}\Gamma_{\xi}\left(  0\right)  .$It describes a stable power-law
solution when $\left\{  -\sqrt{3}<\lambda<0~,~\Gamma_{\xi}\left(  0\right)
<\frac{\lambda}{\sqrt{2}}\right\}  $ or $\left\{  0<\lambda<\sqrt{3}%
~,~\Gamma_{\xi}\left(  0\right)  >\frac{\lambda}{\sqrt{2}}\right\}  $.

f. Point $\bar{B}_{5}$ describes a stable de Sitter universe when $\left\vert
\Gamma_{\xi}^{\prime}\left(  \xi_{0}\right)  \right\vert <\frac{\lambda}%
{\sqrt{2}}.$

g. Point $\bar{B}_{6}$ with coordinates $\bar{B}_{6}=\left(  \bar{A}_{4}%
,\xi_{0}\right)  $ with $\xi_{0}=0$ or $\Gamma_{\xi}\left(  \xi_{0}\right)
=0$. Because of the nonlinearity of the eigenvalues, the regions of the
parameter space, $\left\{  \lambda,\xi_{0}\right\}  $, for which point
$\bar{B}_{6}$ is an attractor, are shown in Fig. \ref{fig9}.

\subsection{Family C}

\begin{figure}[ptb]
\includegraphics[width=0.6\textwidth]{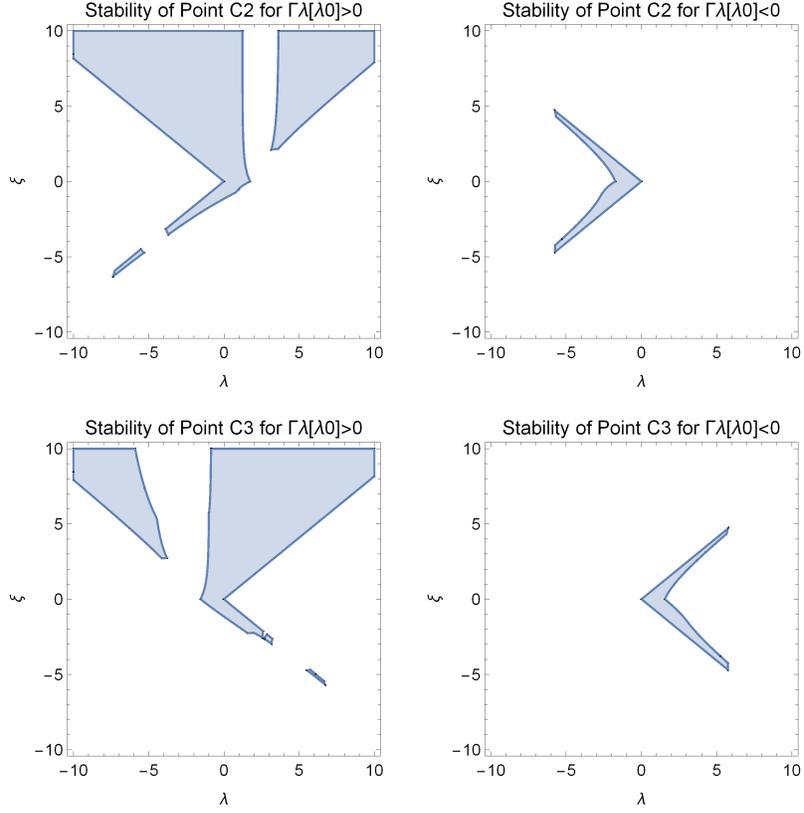} \caption{Region plots for the
variables $\left\{  \lambda,\xi\left(  \lambda\right)  \right\}  $ in which
points $\bar{C}_{2}$ and $\bar{C}_{3}$ are attractors. Left-hand figures are
for$~\Gamma_{\lambda}^{\prime}\left(  \lambda_{0}\right)  >0$ while right-hand
figures are for $\Gamma_{\lambda}^{\prime}\left(  \lambda_{0}\right)  <0$. }%
\label{fig10}%
\end{figure}

We complete our analysis with the third family of critical points which
correspond to the case where $\lambda_{,\phi}\neq const$ and $\xi=\xi\left(
\lambda\right)  $. Using the dynamical system (\ref{pot.12}), (\ref{pot.13}),
(\ref{pot.14}) we find the following critical points $\left(  x_{p}%
,y_{p},\lambda_{p}\right)  $:

a. Point $\bar{C}_{0}=\left(  \bar{A}_{0},\lambda\right)  $ where $\lambda$ is
arbitrary. This point describes a line of points in the space $\left\{
x,y,\lambda\right\}  $. The eigenvalues of the linearized system are
$e_{1}\left(  \bar{C}_{0}\right)  =-\frac{3}{2}~,~e_{2}\left(  \bar{C}%
_{0}\right)  =\frac{3}{2}$ and $e_{3}\left(  \bar{C}_{0}\right)  =0$, from
which we infer that the current point is a source \textquotedblleft
line\textquotedblright.

b. Points $\bar{C}_{1\left(  \pm\right)  }=\left(  \bar{A}_{1\left(
\pm\right)  },\lambda_{0}\right)  $ with $\lambda_{0}=0$ or $\Gamma_{\lambda
}\left(  \lambda_{0}\right)  =1$. These points are always sources, because
they always have a positive eigenvalue. Indeed, the corresponding eigenvalues
are $e_{1}\left(  \bar{C}_{1\left(  \pm\right)  }\right)  =3~,~e_{2}\left(
\bar{C}_{1\left(  \pm\right)  }\right)  =3+\sqrt{\frac{3}{2}}\lambda_{0}$ and
$e_{3}\left(  \bar{C}_{1\left(  \pm\right)  }\right)  =\sqrt{6}\lambda
^{2}\Gamma_{\lambda}^{\prime}\left(  \lambda_{0}\right)  $.

c. Points $\bar{C}_{2,3}$ describes a power-law solution with $\Omega
_{m}\left(  \bar{C}_{2,3}\right)  =0$. Due to the nonlinearity of the
eigenvalues in Fig. \ref{fig10} we present the region of the parameters
$\left\{  \lambda_{0},\xi\left(  \lambda_{0}\right)  \right\}  $ for which
$\bar{C}_{2,3}$ are stable.


d. Point $\bar{C}_{4}$ describes a de Sitter universe and the eigenvalues for
the linearized system are $e_{1}\left(  \bar{C}_{4}\right)  =-3~,~e_{2}\left(
\bar{C}_{4}\right)  =-3$ and $e_{3}\left(  \bar{C}_{4}\right)  =0$. Since
$e_{3}=0$ means that we need to use central manifold theorem: it implies that
$\bar{C}_{4}$ is always a future attractor of the dynamical system.

e. Point $\bar{C}_{5}$ with eigenvalues $e_{1}\left(  \bar{C}_{5}\right)
=-3$,~$e_{2,3}\left(  \bar{C}_{5}\right)  =-\frac{3}{2}\left(  1\pm
\sqrt{1+\left(  \Gamma_{\lambda}\left(  0\right)  -1\right)  \left(  \sqrt
{6}\right)  \xi^{\prime}-2}\right)  $ describes a stable de Sitter solution
when $\left\{  \Gamma_{\lambda}\left(  0\right)  <1,\xi^{\prime}>\sqrt
{\frac{2}{3}}\right\}  $ or~$\left\{  \Gamma_{\lambda}\left(  0\right)
>1,\xi^{\prime}<\sqrt{\frac{2}{3}}\right\}  $.

f. Point $\bar{C}_{6}=\left(  \bar{A}_{4},\lambda_{0}\right)  \,\ $with
$\lambda_{0}=0$ or $\Gamma\left(  \lambda_{0}\right)  =0$, is stable when
$\left\{  \sqrt{\frac{3}{2}}<\lambda~,~\xi>\frac{6-2\lambda^{2}}%
{\sqrt{6\lambda^{2}-9}}\right\}  $ or $\left\{  \lambda<\frac{\sqrt{6}}%
{2}~,~\xi<\frac{6-2\lambda^{2}}{\sqrt{6\lambda^{2}-9}}\right\}  $.

Overall, from the aforementioned analysis, we conclude that in the case of
matter there are two possible de Sitter solutions which can act as attractors
for the expansion dynamics.

\section{Exact solutions}

In the above analysis we have shown that the evolution of the dimensionless
dynamical system is always in a three-dimensional phase space, and there is an
extra free function which must satisfy constraints in order for the critical
point, that is, the solution at the critical point, to exist. In order to
understand this more fully let us consider the field equations (\ref{pot.08}%
)-(\ref{pot.09}) and assume that there is no matter source other than the
scalar field, i.e. $\rho_{m}=p_{m}=0$.

From system (\ref{pot.08})-(\ref{pot.09}) we find%
\begin{equation}
U\left(  \phi\right)  =\frac{\theta^{2}}{3}-\frac{\dot{\phi}^{2}}%
{2}~,~Y\left(  \phi\right)  =-\frac{2}{3}\frac{\dot{\theta}}{\dot{\phi}}%
-3\dot{\phi}.
\end{equation}

Assume now that $\theta=\theta_{0}t^{-1},~$which describes a perfect fluid
solution, and for $\phi=t^{-1}$ we find that%
\begin{equation}
U\left(  \phi\right)  =\frac{\phi^{2}}{2}\left(  \frac{2}{3}\theta_{0}%
^{2}-\phi^{2}\right)  ~,~Y\left(  \phi\right)  =\phi^{2}-\frac{2}{3}\theta
_{0}.
\end{equation}
Hence, this solution is described by a point in family C. In order to study
the stability of the solution we substitute in (\ref{pot.08})-(\ref{pot.09})
$\theta=\theta_{0}t^{-1}+\varepsilon\Theta\left(  t\right)  ,~\phi
=t^{-1}+\varepsilon\Phi\left(  t\right)  $ for the latter potentials and the
solution of the linearized system reveals that functions $\Theta$ and $\Phi$
decay when$~\theta_{0}>3$.

However, this it is not the unique power-law solution as is the case in GR.
Indeed, by selecting the same expansion rate $\theta=\theta t^{-1}$, but now
$\phi=\ln t^{\frac{1}{\alpha}}$, we find
\begin{equation}
U\left(  \phi\right)  =\left(  \frac{\theta_{0}^{2}}{3}-\frac{1}{2\alpha^{2}%
}\right)  e^{-2\alpha\phi}~,~Y\left(  \phi\right)  =\left(  \frac{2}{3}%
\alpha\theta-\frac{1}{\alpha}\right)  e^{-\alpha\phi},
\end{equation}
and this is a solution that is now described by a critical point of family B. \

Moreover, when $\theta=\frac{2}{3}\coth\left(  \theta_{0}\tau\right)  $ and
$\phi=\frac{1}{\alpha t}$ we find
\begin{equation}
U\left(  \phi\right)  =\frac{4}{27}\coth^{2}\left(  \alpha\theta_{0}%
\phi\right)  -\frac{1}{2\alpha^{2}}~,~Y\left(  \phi\right)  =-\frac{8}%
{27}\alpha\theta_{0}\frac{\cosh\left(  \alpha\theta_{0}\phi\right)  }%
{\sinh^{3}\left(  \alpha\theta_{0}\phi\right)  }.
\end{equation}
In a similar way we can construct other kinds of solutions for the scalar
field with the aether field and other kinds of matter sources. For instance,
the latter solution is that of GR with cosmological constant term and a
perfect fluid source.

\section{Conclusions}

\label{sec6}

In this paper we have performed a detailed analysis of the dynamical evolution
of an Einstein-Aether scalar field cosmology in the framework of a
spatially-flat FLRW background universe, where the scalar field is coupled to
the aether field. The model that we analysed depends on two unknown functions,
the first, $U\left(  \phi\right)  ,$ corresponds to the scalar field
potential, while the second function $Y\left(  \phi\right)  $ defines the
coupling between the scalar field and the aether field. The Friedmann equation
is the same with that of Einstein's GR, while the Raychaudhuri acceleration
equation is modified, since the effective pressure term of the scalar field
fluid differs from that in GR.

We use expansion-normalised variables to rewrite the field equations as a
system of algebraic-differential equations, which contains at most three
independent variables. The possible critical points correspond to three
different families of solutions. In family A the scalar field potential is
exponential $U\left(  \phi\right)  =U_{0}e^{\lambda\phi}~$while the coupling
function is given by $~Y\left(  \phi\right)  =Y_{0}+Y_{1}e^{-\frac{\lambda}%
{2}\phi}$. Family B corresponds to the exponential potential for $U\left(
\phi\right)  ,$ while $Y\left(  \phi\right)  $ is arbitrary, while in family
$C$ the scalar field potential $U\left(  \phi\right)  $ is arbitrary.

When we assume there is no other fluid source in the universe, we find that
family A admits four-critical points, while families B and C admit six
critical points. On the other hand, when a pressureless matter source is
introduced, the maximum number of possible critical points is increased by two
for the three families. At this point it is important to mention that family A
corresponds to a specific case of the function $V\left(  \phi,\theta\right)  $
which is determined in \cite{Barrow} and for which the the field equations
admit exact power-law solutions. Moreover, we found that in family B,
power-law solutions exist only when $Y\left(  \phi\right)  $ is approximated
locally by the exponential function $Y\left(  \phi\right)  \simeq
e^{-\frac{\lambda}{2}\phi}$. On the other hand, we found that for family $C$
there is a critical point which describes a de Sitter universe as a future
attractor. Additionally, the critical points of families A and C reduce to
solutions of GR when $Y\left(  \phi\right)  \ $is constant.

Finally, in order to demonstrate the evolution of the dynamical system, we
presented some phase-space diagrams for specific cases as well as a
qualitative evolution for the respective cosmological parameters. From the
latter it is clear that this specific scalar field type of cosmology approach
can describe some key epochs in cosmological evolution.

\begin{acknowledgments}
GP is supported by the scholarship of the Hellenic Foundation for Research and
Innovation (ELIDEK grant No. 633). SB acknowledges support by the Research
Center for Astronomy of the Academy of Athens in the context of the program
\textquotedblleft\textit{Tracing the Cosmic Acceleration}. JDB is supported by
the Science and Technology Facilities Council (STFC) of the United Kingdom.
\end{acknowledgments}

\end{document}